\documentclass[aps,prd,showpacs,groupedaddress,nofootinbib,twocolumn,amsmath,amsfonts,amssymb]{revtex4} 

\usepackage{amsmath}
\usepackage{amssymb}
\usepackage{epsfig}
\usepackage{natbib}

\usepackage{graphicx}
\usepackage{stmaryrd}
\usepackage{bm}
\usepackage{color}
\usepackage{textcomp}

\newcommand{\eps}  {\epsilon}

\newcommand{\diag}{\mathop{\mathrm{diag}}}

\newcommand{\AddrGRAPPA}{GRAPPA Institute, University of Amsterdam, Science Park 904, 1098 XH Amsterdam, Netherlands}
\newcommand{\AddrMPP}{Max-Planck-Institut f\"ur Physik (Werner Heisenberg
Institut),
F\"ohringer Ring 6, 
80805 M\"unchen, Germany}

\begin{document}

\title{Imprints of  {\em CP} violation induced by sterile neutrinos in T2K data}
 
\author{N. Klop}
\thanks{Now at \AddrGRAPPA}
\author{A. Palazzo}
%\thanks{palazzo@mpp.mpg.de}
\affiliation{\AddrMPP}

%\date{\today}

\begin{abstract}

We investigate the impact of light ($\sim$ eV) sterile neutrinos in the long-baseline experiment T2K.
We show that, within the 3+1 scheme, for mass-mixing parameters suggested by the short-baseline anomalies, the interference among the sterile and the atmospheric oscillation frequencies induces a new 
term in the  $\nu_\mu \to \nu_e$ transition probability, which has the same order of magnitude of the standard 3-flavor solar-atmospheric interference term. We show, for the first time, that current T2K data, taken together with the results of the $\theta_{13}$-dedicated reactor experiments, are sensitive to two of the three {\em CP}-violating phases involved in the 3+1 scheme. Both the standard {\em CP}-phase and the new one  ($\delta_{13} \equiv \delta$ and $\delta_{14}$ in our parameterization choice) tend to have a common best-fit value $\delta_{13} \simeq \delta_{14} \simeq -\pi/2$. Quite intriguingly, the inclusion of sterile neutrino effects leads to better agreement between the two estimates of $\theta_{13}$ obtained, respectively, from reactors and T2K, which in the 3-flavor framework are slightly different.

\end{abstract}


\maketitle

\section{Introduction}

Neutrino physics is entering a new era. The discovery of a relatively large value of the long-sought third mixing angle  $\theta_{13}$ has raised hopes of completing the picture of the standard 3-flavor framework. 
 The determination of the two missing unknown properties, i.e., the amount (if any) of leptonic {\em CP}-violation (CPV) and the neutrino mass hierarchy (NMH) are now realistic targets.

Leptonic CPV is a genuine 3-flavor phenomenon~\cite{Cabibbo:1977nk}, which can occur only if no pair of neutrino mass eigenstates is degenerate ($m^2_i - m^2_j \ne 0$ for $i \ne j,~ i,j =1,2,3 $) and if all the three mixing angles ($\theta_{12}, \theta_{23}, \theta_{13}$)  are nonzero. Now that all these (six) necessary conditions are known to be realized in nature, the next task is to ascertain if a further (last) condition is fulfilled, i.e., if the lepton mixing matrix is non-real, or, equivalently, if the {\em CP}-phase $\delta$ is different from $0$ and $\pi$.
The {\em CP}-phase $\delta$ is already being probed by the long-baseline (LBL) accelerator  experiment T2K~\cite{Abe:2013hdq} (and also by MINOS~\cite{Adamson:2014vgd} with less statistical power) in combination with
the reactor $\theta_{13}$-dedicated experiments~\cite{DayaBay_Nu2014, An:2013zwz,Abe:2014bwa,RENO_Nu2014}, which ``fix'' $\theta_{13}$ independently  of $\delta$. Some (weaker) information on such a fundamental phase  is also provided by atmospheric neutrinos~\cite{Suzuki_NIAPP}.
Quite intriguingly, all the latest global analyses~\cite{Capozzi:2013csa,Gonzalez-Garcia:2014bfa,Forero:2014bxa} show a weak indication (close to the 90\% C.L.) of CPV, the best-fit value of the {\em CP}-phase being $\delta \sim -\pi/2$.

An apparently unrelated issue in present-day neutrino physics  is provided by the  
hints of light ($\sim\rm{eV}$) sterile species suggested by the short-baseline (SBL) anomalies (see~\cite{Palazzo:2013me,Kopp:2013vaa,Giunti:2013aea} for reviews).
In the presence of sterile neutrinos, the 3-flavor scheme must be enlarged so as
to include one (or more) mass eigenstates having nonzero admixture with the active flavors.
In such more general frameworks, new {\em CP}-phases automatically appear and, thus, the question naturally
arises as to whether the current and planned LBL experiments, designed to underpin the standard {\em CP}-phase $\delta$, also have some chance to detect the new potential sources of CPV.%
%%%%%%%%%%%%%%%%%%%%%%%%%%%%%%%%%%%%%%%%%%%
\footnote{Previous work on the effects of light sterile neutrinos in LBL setups
can be found in~\cite{Donini:2001xy,Donini:2001xp, Donini:2007yf,Dighe:2007uf,Donini:2008wz,Yasuda:2010rj,Meloni:2010zr,Bhattacharya:2011ee,Donini:2012tt, Hollander:2014iha}.}
%%%%%%%%%%%%%%%%%%%%%%%%%%%%%%%%%%%%%%%%%%

In this work we show, for the first time, that the existing measurements of $\nu_\mu \to \nu_e$ appearance performed
by the LBL experiment T2K,  taken in combination with those of $\bar\nu_e \to \bar\nu_e$ disappearance
deriving from the $\theta_{13}$-dedicated reactor experiments,  are {\em already} able to provide information on the nonstandard sterile-induced {\em CP} phases. In fact, differently from the SBL experiments,
in LBL setups the oscillations induced by the new sterile neutrino species can interfere with
those driven by the two standard squared-mass splittings giving rise to observable effects. In particular,
it turns out that the interference among the sterile and the atmospheric oscillation frequencies induces a new 
term in the  $\nu_\mu \to \nu_e$ transition probability, which has the same order of magnitude of
the standard 3-flavor solar-atmospheric interference term.
 
Working within the simple 3+1 scheme, 
we show that, for mass-mixing parameters suggested by the current SBL fits~\cite{Kopp:2013vaa,Giunti:2013aea}, it is possible to extract quantitative information on two of the three {\em CP} phases involved in the model
(one of them being the standard phase $\delta$). Quite intriguingly, the statistical significance
of the information obtained on the new {\em CP} phase is similar to that achieved for the standard
phase $\delta$. In addition, our analysis evidences that the presence of 4-flavor effects tends to resolve
the slight tension (present within the 3-flavor framework) between the two estimates of $\theta_{13}$ 
extracted, respectively, from T2K and from reactor experiments. 

The rest of the paper is organized as follows. In Sec.~II we introduce the theoretical framework needed 
to discuss the analytical behavior of the LBL $\nu_\mu \to \nu_e$ transition probability in vacuum. 
In Sec.~III we  present the results of the numerical analysis (which includes the matter effects).
In Sec.~IV we draw our conclusions. The paper is closed by an appendix dedicated to the analytical 
treatment of the MSW effects relevant for  the LBL setups within the 3+1 scheme.

\section{Theoretical Framework}

Light sterile neutrinos are introduced in the so-called $3+N_s$ schemes, where the 
$N_s$ new mass eigenstates are assumed to be separated from the three standard ones by large splittings,
giving rise to the hierarchal pattern $\Delta m^2_{12} \ll |\Delta m^2_{13}| \ll |\Delta m^2_{1k}|\,\, 
(k = 4, ..., 3+N_s)$, with $\Delta m^2_{ij} \equiv m^2_j -m^2_i$. This implies that the fast
oscillations induced by the new squared-mass  splittings are completely averaged in all setups 
sensitive to  the solar squared-mass difference ($\Delta m^2_{12}$)   and the atmospheric one
 ($\Delta m^2_{13}$). In this work, for definiteness, we consider the simplest 3+1 scheme.

In the presence of a fourth sterile
neutrino $\nu_s$, the flavor ($\nu_\alpha$, $\alpha = e, \mu, \tau, s$) and the mass eigenstates 
($\nu_i, i =1,2,3,4$) are connected through a $4\times4$ unitary mixing matrix $U$, which depends
on six complex parameters~\cite{Schechter:1980gr}. Such a matrix can  be expressed as the 
product  of six complex elementary rotations,  which define six real mixing angles and six {\em CP}-violating phases.
Of the six phases three are of the Majorana type and are unobservable in oscillation processes, while the three remaining ones are of the Dirac type. As it will appear clear in what follows, for the treatment  of the transitions involved in LBL setups, a particularly convenient choice of the parameterization of the mixing matrix is%
%%%%%%%%%%%%%%%%%%%%%%%%%%%%%%%%%%%%%%%%%%%%%%%%%%%%%%%%%%%%%%%%
\footnote{An equally convenient choice is obtained by associating one of the new {\em CP} phases 
to $R_{24}$ instead of $R_{14}$, i.e. by replacing the product  $R_{24} \tilde R_{14}$ with  $\tilde R_{24} R_{14}$.} 
%%%%%%%%%%%%%%%%%%%%%%%%%%%%%%%%%%%%%%%%%%%%%%%%%%%%%%%%%%%%%%%%
%--------------------------------------------------------------------------------------------------------------------------------------
\begin{equation}
\label{eq:U}
U =   \tilde R_{34}  R_{24} \tilde R_{14} R_{23} \tilde R_{13} R_{12} %\equiv  \bar U U_{3\nu}\,, 
\end{equation} 
%--------------------------------------------------------------------------------------------------------------------------------------
where $R_{ij}$ ($\tilde R_{ij}$) represents a real (complex) $4\times4$ rotation in the ($i,j$) plane
containing the $2\times2$ submatrix 
%--------------------------------------------------------------------------------------------------------------------------------------
\begin{eqnarray}
\label{eq:R_ij_2dim}
     R^{2\times2}_{ij} =
    \begin{pmatrix}
         c_{ij} &  s_{ij}  \\
         - s_{ij}  &  c_{ij}
    \end{pmatrix}
\,\,\,\,\,\,\,   
     \tilde R^{2\times2}_{ij} =
    \begin{pmatrix}
         c_{ij} &  \tilde s_{ij}  \\
         - \tilde s_{ij}^*  &  c_{ij}
    \end{pmatrix}
\,,    
\end{eqnarray}
%--------------------------------------------------------------------------------------------------------------------------------------
in the  $(i,j)$ subblock, with
%--------------------------------------------------------------------------------------------------------------------------------------
\begin{eqnarray}
 c_{ij} \equiv \cos \theta_{ij} \qquad s_{ij} \equiv \sin \theta_{ij}\qquad  \tilde s_{ij} \equiv s_{ij} e^{-i\delta_{ij}}.
\end{eqnarray}
%--------------------------------------------------------------------------------------------------------------------------------------
The  parameterization in Eq.~(\ref{eq:U}) has the following properties: (i) For vanishing mixing
involving the fourth state $(\theta_{14} = \theta_{24} = \theta_{34} =0)$ 
it reduces to the 3-flavor mixing matrix in its standard parameterization~\cite{PDG}
with the identification $\delta_{13} \equiv \delta$. 
(ii) The leftmost positioning of the matrix $\tilde R_{34}$ allows us to eliminate 
the mixing angle $\theta_{34}$ (and the related {\em CP} phase $\delta_{34}$) from the expression 
of the $\nu_{\mu} \to \nu_{e}$ conversion probability in vacuum. In matter, the transition probability
depends also on $\theta_{34}$ and $\delta_{34}$. However, the sterile-induced matter
perturbations are extremely small in T2K and such dependency is completely negligible (see the Appendix). 
(iii)  For small values of $\theta_{13}$ and of the mixing angles involving the fourth mass eigenstate, one has $|U_{e3}|^2 \simeq s^2_{13}$, $|U_{e4}|^2 = s^2_{14}$ (exact), $|U_{\mu4}|^2  \simeq s^2_{24}$ and $|U_{\tau4}|^2 \simeq s^2_{34}$, with a clear physical interpretation of the new mixing angles. 

Before considering the 4-flavor transition probability relevant for T2K,
we recall a basic property of the 3-neutrino framework, which will be helpful
in understanding the more general 4-neutrino case.  In the presence of CPV, 
at LBL experiments one expects a nonzero value of the asymmetry
%--------------------------------------------------------------------------------------------------------------------------------------
\begin{equation}
\label{eq:Asy_1}
A_{\mu e}^{\mathrm {CP}} \equiv P(\nu_\mu \to \nu_e) - P(\bar\nu_\mu \to \bar\nu_e)\,,
\end{equation}
%--------------------------------------------------------------------------------------------------------------------------------------
which, in the 3-flavor (vacuum) case can be expressed as
%.........................................................................................................................................
\begin{equation}
\label{eq:Asy_1}
A_{\mu e}^{\mathrm {CP}} = 16\, J \sin \Delta_{12} \sin \Delta_{23} \sin \Delta_{31}\,,  
\end{equation}
%.........................................................................................................................................
where $\Delta_{ij} \equiv  \Delta m^2_{ij}L/4E$ (L being the baseline and E the neutrino energy)
and $J$ is the Jarlskog invariant~\cite{Jarlskog:1985ht}
%.........................................................................................................................................
\begin{equation}
J = {\rm {Im}} [U_{e 2} U_{\mu 3}  U_{e 3}^* U_{\mu 2}^*]\,.   
\end{equation}
%.........................................................................................................................................
The mass pattern chosen by nature for the three mass eigenstates 
is such that in the LBL setups we have $\Delta \equiv \Delta_{13} \simeq  \Delta_{23} \sim O(1)$
and  $\Delta_{12} = \alpha \Delta$, with%
%%%%%%%%%%%%%%%%%%%%%%%%%%%%%%%%%%%%%%%%%%%%%%%%%%%%%%%%%%%%%%%%
\footnote{Note that the sign of both $\Delta \equiv \Delta_{13}$ and $\alpha \equiv \Delta_{12} /\Delta_{13}$ changes 
with a swap of the mass hierarchy $(\Delta_{13} \to - \Delta_{13})$, while the sign of the product $\alpha \Delta$ is invariant (positive).} 
%%%%%%%%%%%%%%%%%%%%%%%%%%%%%%%%%%%%%%%%%%%%%%%%%%%%%%%%%%%%%%%% $|\alpha| 
$|\alpha| \simeq 0.03$. Consider,
however, an ideal case where in the same setups we have $|\Delta| \equiv |\Delta_{13}|\simeq |\Delta_{23}| \gg1$,
while leaving $\Delta_{12}$ unaltered (this would correspond to a much larger value of $|\Delta m^2_{13}|$).
In this situation, the average over the finite energy resolution of the experiment would wash out
 the fast oscillating terms in Eq.~(\ref{eq:Asy_1}), giving rise to
 %.........................................................................................................................................
\begin{equation}
- \langle \sin \Delta_{23} \sin \Delta_{31} \rangle \simeq  \langle \sin^2\Delta \rangle = 1/2\,, 
\end{equation}
%.........................................................................................................................................
thus obtaining for the {\em CP} asymmetry
%.........................................................................................................................................
\begin{equation}
\langle A_{\mu e}^{\mathrm {CP}}\rangle = -8\, J \sin \Delta_{12}. 
\end{equation}
%.........................................................................................................................................
This ideal case shows that CPV is not erased by the average 
over the two large frequencies. CPV cancellation would occur only if 
the third frequency were also very large ($\Delta_{12} \gg1$).  
Consider now a realistic  3+1 scheme.  In this case, the oscillations induced by the
three new large oscillation frequencies  $|\Delta_{new}|  \equiv |\Delta_{14}| \simeq |\Delta_{24}| \simeq |\Delta_{34}| \gg 1$ 
will get averaged, and the information on their values will be lost.
On the other hand, CPV effects will survive, since there are other two
finite frequencies, the atmospheric [$|\Delta| \equiv |\Delta_{13}|\simeq |\Delta_{23}| \sim O(1)$]
and the solar one ($\Delta_{12} = \alpha \Delta$). Therefore, it will be natural to expect
a dependency on the  {\em CP}-phases in the flavor conversion process.  

Let us now come to the expression of the $\nu_\mu \to \nu_e$ transition probability
probed in T2K.  We recall that in the 3-flavor
vacuum case, such a probability can be written as the sum of three terms
%.........................................................................................................................................
\begin{equation}
P^{3\nu}_{\mu e}  = P^{\rm {ATM}}+ P^{\rm {SOL}}  + P^{\rm {INT}}\,,  
\end{equation}
%.........................................................................................................................................
where the first two terms are positive definite probabilities induced by
the atmospheric and the solar squared-mass splittings, while the third 
is generated by their interference and can assume both positive and negative values.
An expansion in  the small parameters  $s_{13} \simeq 0.15$ and 
$\alpha \simeq \pm0.03$ (supposed to have the same order $\epsilon$) 
provides, at the second order in $\epsilon$, the well-known expressions~\cite{Cervera:2000kp}
%.........................................................................................................................................
\begin{eqnarray}
\label{eq:Pme_3nu_vac_atm}
 &\!\! \!\! \!\! \!\! \!\! \!\! \!\!  P^{\rm {ATM}}_{3\nu} &\!\! \simeq\,  4 s_{23}^2 s^2_{13}  \sin^2{\Delta}\,,\\
 \label{eq:Pme_3nu_vac_sol}
 &\!\! \!\! \!\! \!\! \!\! \!\! \!\! \!\! P^{\rm { SOL}}_{3\nu} &\!\!  \simeq\,   4 c_{12}^2 c_{23}^2 s_{12}^2 (\alpha \Delta)^2\,,\\
 \label{eq:Pme_3nu_vac_int}
 &\!\! \!\! \!\! \!\! \!\! \!\! \!\! \!\! P^{\rm {INT}}_{3\nu} &\!\!  \simeq\,  8 s_{13} s_{12} c_{12} s_{23} c_{23} (\alpha \Delta)\sin \Delta \cos({\Delta + \delta_{13}})\,.
\end{eqnarray}
%.........................................................................................................................................
While all these three terms are formally of the same (second) order in $\epsilon$, 
they have quite different sizes, since $|\alpha|$ is much smaller than $s_{13}$.
Around the first oscillation maximum ($\Delta \sim \pi/2$) probed by T2K, the atmospheric term
is $\sim 5 \times 10^{-2}$,  the interference term is $\sim 1.3 \times 10^{-2}$, and the solar term 
is $\sim 1.5 \times 10^{-3}$. Indeed, a different kind of expansion, 
considering $s_{13} \sim \epsilon$  and $\alpha \sim \epsilon^2$, is
more appropriate, as already evidenced in~\cite{Asano:2011nj}.  In this case, an
expansion at the third order gives only the (leading) atmospheric
term ($\sim \epsilon^2$) and the (subleading) interference term ($\sim \epsilon^3$).
At the fourth order one recovers the solar term $\propto (\alpha\Delta)^2$ of 
Eq.~(\ref{eq:Pme_3nu_vac_sol}), and two additional terms~\cite{Asano:2011nj}. 
The first term can be interpreted as a tiny modification
of the atmospheric term $\delta P^{\rm {ATM}} \propto s_{13}^4$, and the second one as a 
very small change of the interference term  $\delta P^{\rm {INT}} \propto s_{13}^2 (\alpha\Delta)$.
All three fourth-order terms, for the T2K baseline, have size  
$\lesssim 2 \times 10^{-3}$ and have negligible impact.

Let us now come to the 4-flavor case. The analyses of the SBL anomalies~\cite{Kopp:2013vaa,Giunti:2013aea}
point towards values of $s_{14}$ and  $s_{24}$ which have sizes very similar to $s_{13}$.
Hence, for the purposes of this work, which is limited to the SBL-preferred
region, it is meaningful to assume that all these parameters have
the same common order $\epsilon$ and to consider $\alpha \sim \epsilon^2$.
After averaging over the fast oscillations induced by the large
frequency $\Delta_{14}$, at the fourth order in $\epsilon$, we find
%.........................................................................................................................................
\begin{eqnarray} 
\label{eq:Pme_4nu_vac}
P^{4\nu}_{\mu e}  & \simeq& (1 - s^2_{14} - s^2_{24}) P_{\mu e}^{3\nu} \\
\nonumber
&+& 4 s_{14} s_{24} s_{13} s_{23} \sin\Delta \sin (\Delta + \delta_{13} - \delta_{14}) \\
\nonumber
&-& 4 s_{14} s_{24} c_{23} s_{12} c_{12} (\alpha \Delta) \sin \delta_{14}\\
\nonumber
&+& 2 s_{14}^2 s^2_{24} \,.
\end{eqnarray} 
%.........................................................................................................................................
The first term is the 3-flavor probability multiplied by the suppression factor
$f = 1 - O (\epsilon^2$).  The second and third terms can be ascribed, respectively, to 
the interference of the atmospheric and solar frequencies with the 
new large frequency, which does not appear in the formulas
since it has been averaged out.  The last term can be interpreted as the averaged 
transition probability in an effective 2-flavor description.  
As might have been expected, 
the transition probability is the sum of six contributions
%.........................................................................................................................................
\begin{eqnarray}
\label{eq:Pme_4nu_6_terms}
P^{4\nu}_{\mu e}  &=&  P^{\rm{ATM}} + P^{\rm {SOL}}  + P^{\rm {STR}}\\
\nonumber
 &+& P^{\rm {INT}}_{\rm I} +   P^{\rm {INT}}_{\rm{II}}  + P^{\rm {INT}}_{\rm{III}}\,.
\end{eqnarray}
%.........................................................................................................................................
The first two terms and the fourth one coincide (apart from the suppression factor
$f$) with the three standard 3-flavor terms in Eqs.~(\ref{eq:Pme_3nu_vac_atm})-(\ref{eq:Pme_3nu_vac_int}).
With $P^{\rm {STR}}$ we have indicated the last term in~(\ref{eq:Pme_4nu_vac}).
The last two contributions are the new interference
terms [the second and third terms in Eq.~(\ref{eq:Pme_4nu_vac})].
Inspection of Eq.~(\ref{eq:Pme_4nu_vac}) reveals that  $P^{\rm{INT}}_{\rm{II}}$
is $O(\eps^3)$ so its size is expected to be comparable to that of the standard inference
term  $P^{\rm {INT}}_{\rm{I}}$ [see Eq.~(\ref{eq:Pme_3nu_vac_int})]. 
Both $P^{\rm {STR}}$ and $P^{\rm {INT}}_{\rm{III}}$ are $O (\eps^4)$
like $P^{\rm {SOL}}$.

Let us come back for a moment to Eq.~(\ref{eq:Pme_4nu_vac}). From this expression
we can observe that the last three terms depend on the product $s_{14} s_{24}$,
and, therefore, they depend on the effective appearance mixing angle, defined as  
%.........................................................................................................................................
\begin{eqnarray}
\label{eq:Appearance_angle}
\sin 2\theta_{\mu e} \equiv 2 |U_{e4}|  |U_{\mu 4}|  \simeq 2 s_{14} s_{24} \,,
\end{eqnarray}
%.........................................................................................................................................
which is  the amplitude probed by the SBL  $\nu_\mu \to \nu_e$ appearance  experiments. 
The first term in Eq.~(\ref{eq:Pme_4nu_vac}) depends on a different combination of the
two mixing angles. However, in the particular case $s^2_{14} = s^2_{24}$,
this combination can be expressed as
%.........................................................................................................................................
\begin{eqnarray}
\label{eq:sum_2_mix}
s_{14}^2 + s_{24}^2  =  2 s_{14}^2 =  2 s_{14} s_{24}  =  \sin 2\theta_{\mu e} \,,
\end{eqnarray}
%.........................................................................................................................................
and the suppression factor $f$ as%
%%%%%%%%%%%%%%%%%%%%%%%%%%%%%%%%%%%%%%%%%%%%%%%
\footnote{It is useful to observe that for a fixed value of the appearance angle
$\sin 2\theta_{\mu e}$, the suppression factor $f$ is always smaller than the one obtained 
in the specific case $s^2_{14} = s^2_{24}$. In fact, the inequality $(s_{14} - s_{24})^2 \ge 0$
implies that $s_{14}^2 + s_{24}^2 \ge 2 s_{14} s_{24} = \sin 2\theta_{\mu e}.$ }
%%%%%%%%%%%%%%%%%%%%%%%%%%%%%%%%%%%%%%%%%%%%%%% 
%.........................................................................................................................................
\begin{eqnarray}
\label{eq:factor}
f = 1 -  \sin 2\theta_{\mu e} \,.
\end{eqnarray}
%.........................................................................................................................................
Therefore, in this particular case, Eq.~(\ref{eq:Pme_4nu_vac}) can be recast  in the form 
%.........................................................................................................................................
\begin{eqnarray}
\label{eq:Pme_app_angle}
P^{4\nu}_{\mu e}  &=& (1 - \sin 2\theta_{\mu e} ) P_{\mu e}^{3\nu} \\
\nonumber
&+& 2  \sin 2\theta_{\mu e} s_{13} s_{23} \sin\Delta \sin (\Delta + \delta_{13} - \delta_{14})\\ 
\nonumber
&-& 2 \sin 2\theta_{\mu e} c_{23} s_{12} c_{12} (\alpha \Delta) \sin \delta_{14}\\
\nonumber
&+& \frac{1}{2}  \sin^2 2\theta_{\mu e}\,,
\end{eqnarray}
%.........................................................................................................................................
in which all the terms [and consequently all six terms in Eq.~(\ref{eq:Pme_4nu_6_terms})]
scale with a definite power of the effective appearance mixing angle. 

%%%%%%%%%%%%%%%%%%%%%%%%%%%%%%%%%%%%%%%%%%%
\begin{figure}[t!]
\vspace*{-0.15cm}
\hspace*{-0.25cm}
\includegraphics[width=9.2 cm]{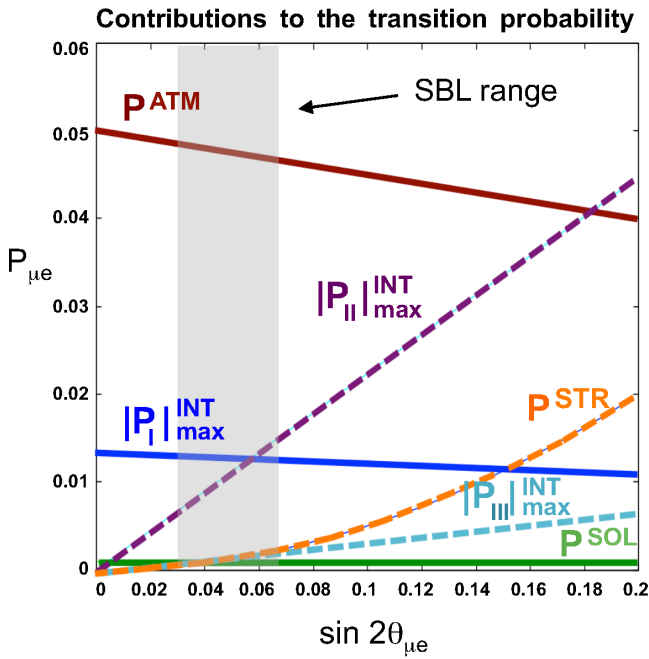}
\vspace*{-0.7cm}
\caption{Behavior of the six contributions to the transition probability in Eq.~(\ref{eq:Pme_4nu_6_terms}) as 
a function of the effective appearance mixing angle $\sin 2\theta_{\mu e}$ in the case  $s_{14}^2 = s_{24}^2$
for $s^2_{13} = 0.025$.  The maximum absolute value is plotted for the three interference terms. The curves 
correspond to the first oscillation maximum $\Delta = \pi/2$ (obtained for $E_\nu \simeq 0.6$\,\,GeV in T2K). 
\label{eq:pme_six_terms}}
\end{figure}  
%%%%%%%%%%%%%%%%%%%%%%%%%%%%%%%%%%%%%%%%%%%%%

As a result, in the particular case $s^2_{14} = s^2_{24}$, the comparison of the terms
involved in the transition probability is straightforward.
Figure~\ref{eq:pme_six_terms} displays (for $\Delta = \pi/2$ and $s^2_{13} = 0.025$)
the behavior of all the six contributions in Eq.~(\ref{eq:Pme_4nu_6_terms})
as a function of $\sin 2\theta_{\mu e}$. We plot the maximum absolute value 
for the three interference terms (which can be positive or negative).
The vertical gray band indicates the range allowed by the SBL anomalies.
As expected from the discussion above, in this range, only the leading 
atmospheric and the two subleading interference terms are relevant,
and the 4-flavor transition probability is approximately given by
%.........................................................................................................................................
\begin{eqnarray}
\label{eq:Pme_4nu_3_terms}
P^{4\nu}_{\mu e}  \simeq  P^{\rm{ATM}} + P^{\rm {INT}}_{\rm I}+   P^{\rm {INT}}_{\rm II}\,.
\end{eqnarray}
%.........................................................................................................................................
Remarkably, the amplitude of the new (atmospheric-sterile) interference term is
almost identical to that of the standard (solar-atmospheric) interference term. 
As a consequence, a big impact is expected on the regions reconstructed by the experiment
T2K for the {\em CP}-phase $\delta_{13}$. More importantly, a similar 
sensitivity to the two {\em CP}-phases $\delta_{13}$ and  $\delta_{14}$ is
expected in the combination of T2K with the reactor experiments. The quantitative 
verification of these qualitative expectations will be the subject of the next section.

 \section{Numerical Analysis}
 
In our numerical analysis we include the reactor experiments Daya Bay and RENO  and the LBL experiment T2K. Concerning T2K the analysis is slightly different for the two cases of 3 and 4 flavors, since in this last case there are appreciable oscillation effects not only at the far detector (FD), Super-Kamiokande, but also at the near detector  (ND), ND280, which must be taken into account. For this reason, we discuss  the two cases separately.  In the calculation of the oscillation probability we have included the MSW effects following the prescriptions described in the Appendix.
 
 \subsection {Treatment of the reactor experiments}  
 
The analysis of the reactor experiments is performed using the total rate information 
and following the approach described in detail in~\cite{Palazzo:2013bsa}.
For both experiments we have used the latest data presented at the Neutrino 2014 
Conference~\cite{DayaBay_Nu2014,RENO_Nu2014} based on 621 live days
(Daya Bay) and 800 live days (RENO). Since the electron antineutrino survival probability
 probed by these experiments is independent of the {\em CP}-violating phases 
 (standard and nonstandard), their estimate of $\theta_{13}$ does not depend on their values. 
We recall that such an estimate is extracted using the ratio of the event rates measured at the far 
and the near sites. Since the fast oscillations induced by $\Delta m^2_{14}$ are averaged out 
at both detector sites, Daya Bay and RENO are not sensitive to 4-flavor effects.
 As a result, their estimate of $\theta_{13}$ is independent of the mixing angle 
 $\theta_{14}$ provided that it is confined to vary in the range we are exploring.%
 %%%%%%%%%%%%%%%%%%%% %%%%%%%%%%%%%%%%%%%%%%%%%%%%%%
 \footnote{It can be shown that corrections to this approximation arise at the order $\epsilon^6$,
 being proportional to $s_{13}^2 s_{14}^4$, and are completely negligible.} 
 %%%%%%%%%%%%%%%%%%%%%%%%%%%%%%%%%%%%%%%%%%%%%%%%%%
 Finally, it must be noted that the estimate of $\theta_{13}$ is essentially identical for the two cases of normal hierarchy
 (NH) and inverted hierarchy (IH).
 
\subsection {Treatment of T2K}
 
 \subsubsection{The 3-flavor case} 
  
We use the T2K results of the $\nu_\mu \to \nu_e$ appearance searches~\cite{Abe:2013hdq},
which reported 28 events with an estimated background of 4.92 events.   
In order to calculate the theoretical expectation for the total number of events
and their binned spectrum in the reconstructed neutrino energy, we convolve the product 
of the $\nu_\mu$ flux~\cite{Abe:2012av}
(tables provided on the T2K home page~\cite{T2K_webpage}),
the charged current quasielastic cross-section (estimated from~\cite{Abe:2013hdq}), and 
the $\nu_\mu \to \nu_e$  transition probability, with an appropriate
energy resolution function, which incorporates the correlation among the true and
the reconstructed neutrino energy. For the energy resolution we have adopted a
Gaussian function with width $\sigma_E = 15\%\, \sqrt E$. We have checked that our prediction for
the binned spectrum of events is in very good agreement with that shown by the collaboration in~\cite{Abe:2013hdq}.
We have performed the analysis using both the total rate information and the full spectrum,
observing very small differences between the two cases. This is due to the limited statistics currently available,
and to the effect of the smearing induced by the energy resolution. As explained in the next subsection, 
in the 4-flavor case we could consistently perform only a total rate analysis. Therefore, for homogeneity,
in the 3-flavor case, we report the results obtained with the total rate information.

\subsubsection{The 4-flavor case}

In the case of 4-flavor oscillations the T2K near detector is sensitive to 
the oscillations induced by the new $\Delta m^2_{14}$, since at the baseline 
$L^{\rm {ND}} = 280~$m, $\Delta_{14}^{\rm {ND}} \sim O(1)$. Regard this, it should be noted  that the published neutrino
 fluxes $\phi_\nu (E)$ (both those relative to ND280 and their extrapolations at Super-Kamiokande) 
 are constrained with the measurements performed at the ND (see~\cite{Abe:2012av}).
 More precisely, the available fluxes are not the 
 original output of the dedicated simulation programs, but are a  ``postfit" version
 of these. Basically, the fluxes are anchored to the ND measurements, compatibly with
 the pulls of the nuisance parameters of the model of the original (``prefit") fluxes. The anchoring
 of the fluxes to the ND measurements introduces an overall normalization factor and
 appreciable energy distortions, the typical size of which is $\sim 10$\%. These effects can be
 appreciated, for example, in Fig.~1 of~\cite{Abe:2013hdq}.
  
 This procedure is designed for the 3-flavor case, in which the oscillation effects at
 the ND are negligible; the procedure in this case improves the estimate of the nonoscillated fluxes.
 In contrast, in the presence of 4-flavor effects, the available postfit fluxes are no more
 an accurate estimate of the nonoscillated fluxes, since they partially incorporate 
 the effects of the 4-flavor oscillations occurred at the ND. In the range of the small mixing
 angles we are exploring in this work, at the ND it is expected an energy dependent suppression
 of the nonoscillated fluxes whose amplitude ($\simeq  4 s_{24}^2 = 0.1$) is modulated
 by the $O(1)$ phase  $\Delta_{14}^{\rm {ND}} = \Delta m^2_{14}L^{\rm {ND}}/4E$. A suppression 
 of this size is certainly allowed by the nuisance parameters of the flux model.%
 %%%%%%%%%%%%%%%%%%%%%%%%%%%%%%%%%%%%%%%%%%%%%%%%%%%%%%%%%
 \footnote{Indeed, the ND spectrum normalization appears to be  
 a few per cent lower than the prefit one, most of the suppression being 
 concentrated in a region close to the peak of the muon momentum
 distribution (see Fig.~1 of~\cite{Abe:2013hdq}).}
 %%%%%%%%%%%%%%%%%%%%%%%%%%%%%%%%%%%%%%%%%%%%%%%%%%%%%%%%%
 Therefore, for the range of parameters under considerations, the postfit fluxes 
 completely incorporate the 4-flavor effects (if these are present). In this situation, it is problematic to perform 
 an accurate 4-flavor spectral analysis from outside the T2K Collaboration, since one would need to model the original fluxes (with
 their uncertainties) and perform a simultaneous fit of both the ND and the FD event spectra
 varying the oscillation parameters $\theta_{24}$ and $\Delta m^2_{14}$.  For these reasons 
 we limit our work to a total rate analysis. 
 
 We have checked that in the range of $\Delta m^2_{14} \in  [0.1 - 10]$~eV$^2$, the total rate suppression 
 at the ND  varies in the range $[2, 8]\%$. For definiteness, in the analysis we assume $\Delta m^2_{14} = 1 $~eV$^2$, 
 for which the total rate suppression is $\sim 4\%$. Consequently, we have increased  the
 normalization of the published $\nu_\mu$ flux by the same amount. Since the FD total rate is proportional
to the product of the nonoscillated flux and the transition probability $P_{\mu e}$, in the fit, a larger
flux will be compensated by a lower $P_{\mu e}$. By observing that the leading 
term  of $P_{\mu e}$ is proportional to $\sin^2 \theta_{13}$,  we can deduce that 
the estimate of $\sin^2\theta_{13}$ will be slightly smaller ($\sim -3\%$) 
with respect to the 3-flavor case, as we have explicitly checked numerically.
Therefore, when interpreting the results of the 4-flavor analysis, one should bear in mind 
that this (small) effect is at play, together with those (larger) described in Sec.~II, which are genuine 
LBL effects. 
 
We remark that the fake energy distortions introduced in the fluxes 
by the ND280 anchoring procedure are also present, basically unaltered,
in the far detector fluxes (which are an extrapolation of the ND fluxes). For this reason, it would be 
pointless (and wrong) to perform a spectral 4-flavor analysis  of the FD data using the published fluxes.
As we have shown in the previous subsection on the 3-flavor analysis,  
the T2K sensitivity is currently dominated by the total rate information.
 Therefore, it is legitimate to expect that the total rate information will give 
accurate results in the 4-flavor case as well, provided that the fake distortions 
are of the order of a few per cent.  Of course, the situation would be different for 
values of $s_{24}^2$ much larger than those considered in this work, in which case 
a dedicated spectral analysis would be inescapable. The same conclusion would hold
if the T2K $\nu_\mu \to \nu_e$ appearance measurement had considerably 
larger statistics.

A final remark is in order concerning the treatment of the atmospheric mixing angle $\theta_{23}$.
It can been shown that the far/near ratio in the $\nu_\mu \to \nu_\mu$ channel is almost unaffected
by 4-flavor effects, as we have explicitly checked numerically. 
Therefore, the estimate of $\theta_{23}$ is expected to be very stable with respect to the 
4-flavor perturbations induced by the small values of $s_{24}^2$ we are considering.
Therefore, in the 4-flavor analysis we have marginalized $\theta_{23}$
taking into account the constraint  $\sin^2\theta_{23} \simeq 0.51\pm 0.055$ 
obtained, within the 3-flavor framework, from the $\nu_\mu \to \nu_\mu$ disappearance measurement~\cite{Abe:2014ugx} 
performed by T2K. This assumption would not be justified for larger values
of $\theta_{24}$, in which case the estimate of $\theta_{23}$ should be obtained
by analyzing the $\nu_\mu \to \nu_\mu$  disappearance results within a 4-flavor framework.

%%%%%%%%%%%%%%%%%%%%%%%%%%%%%%%%%%%%%%%%%%%
\begin{figure}[t!]
\vspace*{-3.01cm}
\hspace*{-0.90cm}
\includegraphics[width=13.0 cm]{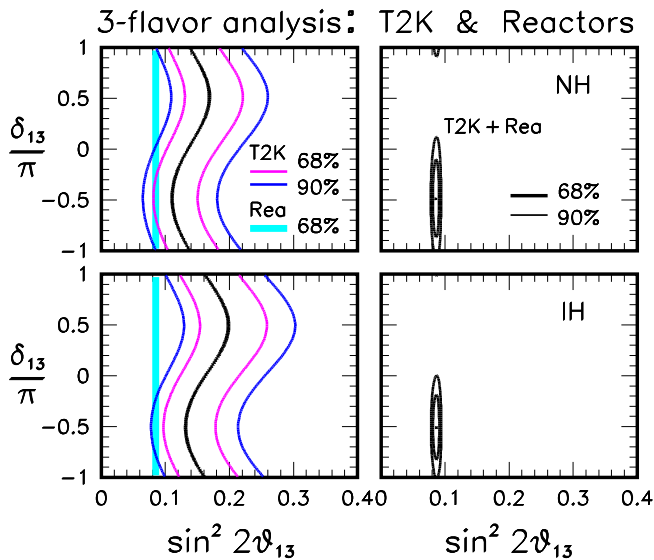}
\vspace*{-3.0cm}
\caption{Left panels: Regions allowed by T2K and by reactor experiments for normal hierarchy (upper panel) and inverted hierarchy (lower panel). Right panels: Regions allowed by their combination. The mixing angle $\theta_{23}$ is marginalized away. The confidence levels refer to 1 d.o.f. ($\Delta \chi^2 = 1.0, 2.71$).
\label{fig:4pan_3nu}}
\vspace*{-0.0cm}
\end{figure}  
%%%%%%%%%%%%%%%%%%%%%%%%%%%%%%%%%%%%%%%%%%%%%  

%%%%%%%%%%%%%%%%%%%%%%%%%%%%%%%%%%%%%%%%%%%
\begin{figure}[b!]
\vspace*{-3.01cm}
\hspace*{-0.90cm}
\includegraphics[width=13.0 cm]{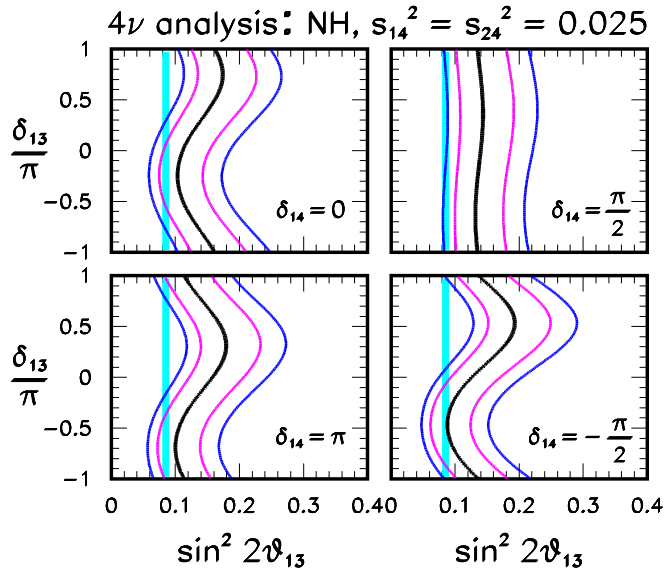}
\vspace*{-2.9cm}
\caption{Regions allowed by T2K for four values of the {\em CP}-phase $\delta_{14}$.  Normal hierarchy is assumed. 
The mixing angle $\theta_{23}$ is marginalized away. The vertical band represents the region allowed by reactor experiments.
Confidence levels are as in Fig.~\ref{fig:4pan_3nu}.
\label{fig:4pan_4nu_nh}}
\end{figure}  
%%%%%%%%%%%%%%%%%%%%%%%%%%%%%%%%%%%%%%%%%%%%%

\subsection{Results of the 3-flavor analysis}

In the 3-flavor analysis, the two mixing angles  ($\theta_{13}$, $\theta_{23}$) 
and the {\em CP}-phase $\delta_{13}$ are treated as free parameters, taking into account the
external prior $\sin^2 \theta_{23} \simeq 0.51 \pm 0.055$ provided by the $\nu_\mu \to \nu_\mu$ 
disappearance measurement~\cite{Abe:2014ugx} performed by T2K. For the atmospheric mass splitting we use the best-fit value
of  $\Delta m^2_{13}$ obtained in the same analysis. The solar
mass-mixing parameters are fixed at the best-fit value obtained in the global analysis~\cite{Capozzi:2013csa}.

Figure~\ref{fig:4pan_3nu} shows the results of the analysis for the two
cases of NH (upper panels) and IH (lower panels) in the plane spanned
by the two variables [$\sin^2 2\theta_{13}, \delta_{13}$], the atmospheric 
mixing angle $\theta_{23}$ having been marginalized away. 
The left panels report the T2K-allowed regions for the confidence levels
68\% and 90\% (1 d.o.f), identical to those used by the T2K Collaboration,
so as to facilitate comparison. Our results are basically superimposable to
those obtained by the collaboration (see Fig.~5 in~\cite{Abe:2013hdq}).
The thin vertical band displayed in both panels represents the range allowed
at 68\% C.L. for $\theta_{13}$ by the reactor experiments. As already noticed in
the global analyses~\cite{Capozzi:2013csa,Gonzalez-Garcia:2014bfa,Forero:2014bxa}
and in partial fits performed by various experimental collaborations, the T2K-allowed
bands lie at values of $\theta_{13}$, which are somewhat larger compared to 
the range identified by reactors. As a result, as evident in the two right panels, the combination 
of the reactor experiments with T2K tends to select values of $\delta \sim -\pi/2$, disfavoring
the cases of no CPV ($\delta_{13} = 0, \pi$) at roughly the 90\% C.L. In addition, a weak
preference for the case of normal hierarchy emerges  ($\chi^2_{\rm {NH}} - \chi^2_{\rm {IH}} \simeq -0.8$).

%%%%%%%%%%%%%%%%%%%%%%%%%%%%%%%%%%%%%%%%%%%
\begin{figure}[b!]
\vspace*{-3.01cm}
\hspace*{-0.90cm}
\includegraphics[width=13.0 cm]{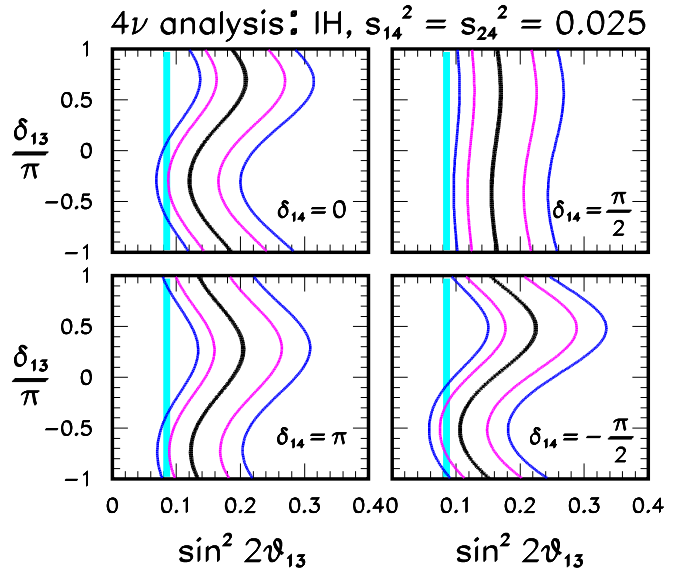}
\vspace*{-2.9cm}
\caption{Regions allowed by T2K for four values of the {\em CP}-phase $\delta_{14}$.  
Inverted hierarchy is assumed. The vertical band represents the region allowed by reactor experiments.
The mixing angle $\theta_{23}$ is marginalized away. Confidence levels are as in Fig.~\ref{fig:4pan_3nu}.
\label{fig:4pan_4nu_ih}}
\end{figure}  
%%%%%%%%%%%%%%%%%%%%%%%%%%%%%%%%%%%%%%%%%%%%% 

\subsection{Results of the 4-flavor analysis}

As discussed in detail in the Appendix,
in the 3+1 scheme, the role of matter effects is very similar to the 3-flavor case.
Basically (in comparison to the vacuum case), they tend to increase (decrease)
the theoretically expected T2K rate in the case of NH (IH), with a consequent downward (upward) 
shift of the range preferred for $\theta_{13}$. The ``wiggle" structure
of the allowed regions is basically the same for the two mass hierarchies (see Figs.~3 and~4).
The regions obtained for the case of IH are essentially shifted towards larger values
of $\theta_{13}$ and slightly expanded with respect to those obtained in the NH case.
 We describe in detail the results only for NH, the interpretation of the IH case
 being straightforward.

%%%%%%%%%%%%%%%%%%%%%%%%%%%%%%%%%%%%%%%%%%%
\begin{figure}[b!]
\vspace*{-2.8cm}
\hspace*{-0.50cm}
\includegraphics[width=12.5 cm]{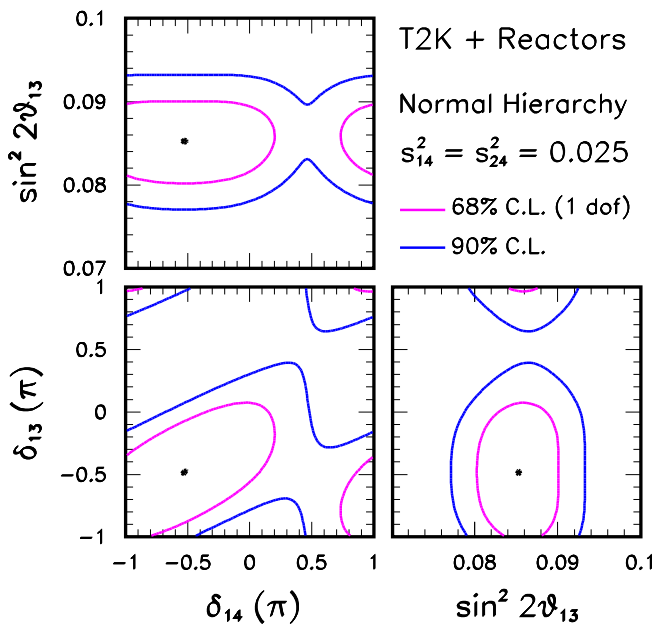}
\vspace*{-1.7cm}
\caption{Regions allowed by the combination of T2K and reactor experiments for the case of normal hierarchy. 
The mixing angle $\theta_{23}$ is marginalized away. 
\label{fig:3pan_4nu_nh}}
\end{figure}  
%%%%%%%%%%%%%%%%%%%%%%%%%%%%%%%%%%%%%%%%%%%%%% 

Figure~\ref{fig:4pan_4nu_nh} displays the results of the 4-flavor analysis for the case
of NH. The four panels represent the T2K-allowed regions in the usual
plane [$\sin^2 2\theta_{13}, \delta_{13}$] for four different choices of the new {\em CP}-phase $\delta_{14}$.
We have fixed the 4-flavor parameters at the following values:
$s_{14}^2 = s_{24}^2 = 0.025$, $s^2_{34} = 0$, $\delta_{34} = 0$ and $\Delta m^2_{14} = 1~$eV$^2$.
As a benchmark we also report the range allowed for $\theta_{13}$ by reactors, which is identical
to the standard case. 
A quick comparison of the four panels of Fig.~\ref{fig:4pan_4nu_nh} with the 3-flavor case
(left upper panel of Fig.~\ref{fig:4pan_3nu})  shows the large impact of the 4-flavor
effects on the structure of the T2K wiggles. The behavior of the curves can be easily 
interpreted, taking into account that the dominant contribution to the total rate comes
from a region of the energies close to the first oscillation maximum,
where  $\Delta \sim \pi/2$. 
Inspection of Eq.~(\ref{eq:Pme_3nu_vac_int}) shows that the standard interference term is
proportional to $-\sin \delta_{13}$.  From Eq.~(\ref{eq:Pme_4nu_vac}) we see that
for $\delta_{14} = \pi/2$, the new interference term is proportional to $\sin \delta_{13}$.
Therefore, in this case the two terms are in opposition of phase and having similar amplitudes
their sum tends to cancel out, making the wiggles almost disappear (right upper panel).
In this case, the T2K region is basically a vertical band.

For $\delta_{14} = -\pi/2$ (right lower panel of Fig.~3) the two inference terms 
have the same phase and the horizontal excursion of the wiggles is increased (roughly doubled).
As a benchmark, the best-fit curve excursion range is $[0.11, 0.17]$ in the 3-flavor case,
while it is $[0.09, 0.19]$ in the 4-flavor one. In the two cases $\delta_{14} = 0, \pi$ (left panels of Fig.~3) 
the new interference term is proportional to $\pm \cos \delta_{13} = \pm \sin (\pi/2 -\delta_{13})$ 
and, thus, it has a $\pm \pi/2$ difference of phase with respect to the standard one.
As a result, in those two cases, the behavior of the T2K bands is intermediate between 
the two cases  $\delta_{14} = (-\pi/2, \pi/2)$.

%%%%%%%%%%%%%%%%%%%%%%%%%%%%%%%%%%%%%%%%%%%%%
\begin{figure}[b!]
\vspace*{-2.8cm}
\hspace*{-0.50cm}
\includegraphics[width=12.5 cm]{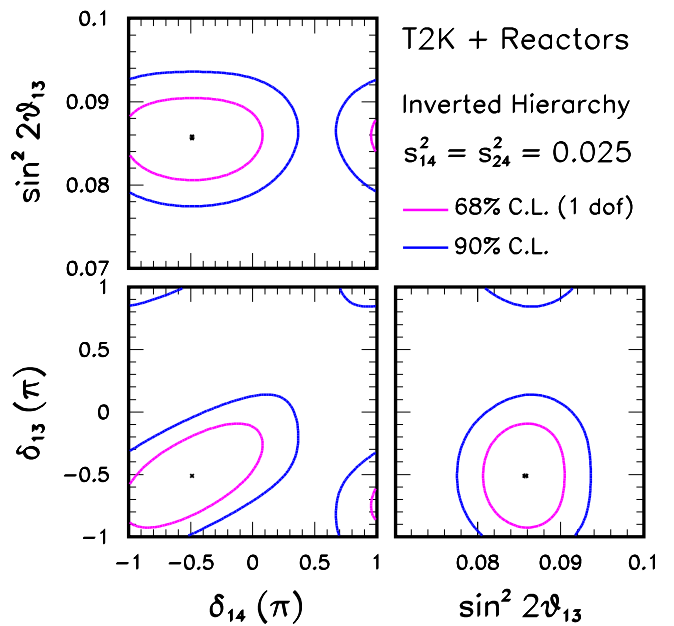}
\vspace*{-1.7cm}
\caption{Regions allowed by the combination of T2K and reactor experiments for the case 
of inverted hierarchy. The mixing angle $\theta_{23}$ is marginalized away. 
\label{fig:3pan_4nu_ih}}
\end{figure}  
%%%%%%%%%%%%%%%%%%%%%%%%%%%%%%%%%%%%%%%%%%%%%%%

It is interesting to note that, in the presence of 4-flavor effects, a better agreement between
the estimates of $\theta_{13}$ from reactors and T2K can be obtained. In particular, this occurs for 
$\delta_{14} \simeq -\pi/2$, which is, therefore, expected to be the best-fit value in
the combined analysis of reactors with T2K. We recall that a
a (small) part of the shift towards lower values of $\theta_{13}$ of the T2K bands
is imputable to the re-normalization of the nonoscillated $\nu_\mu$ flux, which we
have incorporated in our analysis in order to take into account the effect of the oscillations
at the near detector ND280 (see the discussion in subsection III.A.2).

As a last step in our $4$-flavor analysis, we perform the combination of T2K with reactors. 
In this more general analysis, we treat the two mixing angles ($\theta_{13}, \theta_{23}$) and
the two {\em CP}-phases  $(\delta_{13}, \delta_{14})$ as free parameters, while fixing
the remaining 4-flavor parameters at the same values used before:
$s_{14}^2 = s_{24}^2 = 0.025$, $s^2_{34} = 0$, $\delta_{34} = 0$ and $\Delta m^2_{14} = 1~$eV$^2$.
We have checked that the impact of nonzero $\theta_{34}$ (and consequently of
the associated {\em CP}-phase $\delta_{34}$) is negligible, even considering very large values 
of $\theta_{34}$, well beyond the current bounds. Therefore, the results of the analysis, albeit formally obtained for
the fixed value $\theta_{34} = 0$, are indeed more general and are equivalent to those that one would obtain by
treating $\theta_{34}$ and $\delta_{34}$ as free parameters and marginalizing over them. 
The insensitivity to $\theta_{34}$ in vacuum is obvious from the formulas presented in Sec.~II. 
The reason of its irrelevance also in matter is explained in the Appendix.

Similar to the 3-flavor case, in the T2K + reactors combination,
the ({\em CP}-phases-independent) estimate of $\theta_{13}$ 
provided by the reactor experiments selects those subregions of the T2K bands
that have a statistically significant overlap with such an estimate. These, in turn, 
correspond to allowed regions in the plane $[\delta_{13}, \delta_{14}]$ 
spanned by the two {\em CP}-phases. Figures~\ref{fig:3pan_4nu_nh} and~\ref{fig:3pan_4nu_ih} 
display such regions for the two cases of NH and IH, together with the 2-dimensional projections 
that have the mixing angle $\theta_{13}$ as one of the two variables.  As expected a preference 
for values of $\delta_{13} \sim \delta_{14} \sim -\pi/2$ emerges.  The absence of {\em CP} violation
is disfavored at a slightly lower confidence level in comparison with
the 3-flavor case. This is imputable to  the larger freedom allowed by the larger parameter
space available in the 3+1 scheme.

\section{Conclusions and Outlook}

We have investigated the impact of light ($\sim$ eV) sterile neutrinos in the long-baseline experiment T2K.
We have shown that, within the 3+1 scheme, for mass-mixing parameters suggested by the short-baseline anomalies, the new term appearing in the  $\nu_\mu \to \nu_e$ transition probability arising from the interference between the sterile and the atmospheric oscillation frequencies has the same order of magnitude of the standard 3-flavor solar-atmospheric interference term. As a result, the current T2K data (in combination with the $\theta_{13}$-dedicated reactor experiments)  are sensitive to two of the three {\em CP}-violating phases involved in the 3+1 scheme. In particular, we found that both the standard phase and the new one  ($\delta_{13} \equiv \delta$ and $\delta_{14}$ in our parameterization choice) tend to have a common 
best-fit value $\delta_{13} \simeq \delta_{14} \sim -\pi/2$. In addition, quite intriguingly, our analysis shows that the inclusion of sterile neutrino effects leads to better agreement of the two estimates of $\theta_{13}$ obtained, respectively, from T2K and reactor data.

Our results make it evident that T2K and other similar LBL experiments (like MINOS  and NO$\nu$A in the near future) should be routinely included in the global fits involving sterile neutrinos.  Finally, we stress that, in the eventuality of a discovery of  light sterile neutrino species at the upcoming SBL experiments, our findings would represent the first indications on the sterile-induced CPV sources. The LBL accelerator experiments that are already operational, and those planned for the future (LBNE, LBNO, and T2HK), will play a key role in extracting more information on the new CPV sector, the exploration of which has, 
with the present work, just begun.

%%%%%%%%%%%%%%%%%%%%%%%%%%%%%%%%%%%%%%%%%%%%%%%%%%
\section*{APPENDIX: TREATMENT OF THE MSW EFFECT}
%%%%%%%%%%%%%%%%%%%%%%%%%%%%%%%%%%%%%%%%%%%%%%%%%%

The Hamiltonian in the flavor basis can be written as
%------------------------------------------------------------------------------------------------------------------------------------
\begin{equation}
    H = U K U^\dagger + V \,,
    \label{eq:Hf}
\end{equation}
% ------------------------------------------------------------------------------------------------------------------------------------
where $K$  denotes the diagonal matrix containing the wave numbers  
$k_i = m^2_i /2E \, (i=1,2,3,4)$ ($m_i^2$  and $E$ being the neutrino squared-mass  and 
energy respectively), while the matrix $V$ incorporates the position-independent matter MSW 
potential~\cite{Wolfenstein:1977ue, smirnov}.
Barring irrelevant factors proportional to the identity, we can define the
diagonal matrix containing the three relevant wave numbers as
% ------------------------------------------------------------------------------------------------------------------------------------
\begin{equation}
K = \diag(0,\, k_ {12},\, k_{13},\, k_{14})\,
\end{equation}
% ------------------------------------------------------------------------------------------------------------------------------------
and the matrix encoding the matter effects, as  
% ------------------------------------------------------------------------------------------------------------------------------------
\begin{equation}
\label{eq:V_matrix}
  V=\mathrm{diag}(V_{CC},\,0,\,0,\, -V_{NC})\,,
 \end{equation}
% ------------------------------------------------------------------------------------------------------------------------------------
where
% ------------------------------------------------------------------------------------------------------------------------------------
\begin{eqnarray}
V_{CC}
&=&
\sqrt{2} \, G_F \, N_e\,
\label{VCC}
\end{eqnarray}
% ------------------------------------------------------------------------------------------------------------------------------------
is the charged-current interaction potential of the electron neutrinos with 
the background electrons having number density $N_e$, and
%------------------------------------------------------------------------------------------------------------------------------------
\begin{eqnarray}
V_{NC}
&=&
- \frac{1}{2} \sqrt{2} G_F N_n\,
\label{VNC}
\end{eqnarray}
% ------------------------------------------------------------------------------------------------------------------------------------
is the neutral-current interaction potential (common
to all the active neutrino species) with the background 
neutrons having number density $N_n$. For later convenience, we  
also introduce the positive-definite ratio
% ------------------------------------------------------------------------------------------------------------------------------------
\begin{equation}
r = - \frac{V_{NC}} {V_{CC}}  = \frac{1}{2} \frac {N_n}{N_e}\,,
\end{equation}
% ------------------------------------------------------------------------------------------------------------------------------------
which in the Earth  crust is approximately $r \simeq 0.5$.
In order to simplify the treatment of matter effects in LBL experiments it is useful to introduce the new basis
%------------------------------------------------------------------------------------------------------------------------------------
\begin{equation}
\bar \nu = \bar U{^\dag} \nu\,,
\end{equation}
% ------------------------------------------------------------------------------------------------------------------------------------
where
% ------------------------------------------------------------------------------------------------------------------------------------
\begin{equation}
\bar U = \tilde R_{34} R_{24} \tilde R_{14}\,
\end{equation}
% ------------------------------------------------------------------------------------------------------------------------------------
is the part of the mixing matrix defined in Eq.~(\ref{eq:U}) containing only the rotations involving the fourth 
neutrino mass state. In this new basis, the Hamiltonian assumes the form
% ------------------------------------------------------------------------------------------------------------------------------------
\begin{equation} \label{eq:Hbar}
   \bar H = \bar H^{\rm {kin}} + \bar H^{\rm {dyn}} = U_{3\nu} K U_{3\nu}^\dagger
    + {\bar U}^\dagger V {\bar U} \,,
\end{equation}
% ------------------------------------------------------------------------------------------------------------------------------------
where the first term is the  kinematic contribution describing the
oscillations in vacuum, and the second one represents a nonstandard dynamical term.
Since $|k_{14}|$ is much bigger than $k_{12}$ and $|k_{13}|$  and much bigger than 1,
we can reduce the dynamics to that of an effective 3-flavor system. Indeed, 
from Eq.~(\ref{eq:Hbar}) one has that the (4,4) entry of $\bar H$ is much bigger than all the other
elements and
(the absolute value of) the fourth eigenvalue of  $\bar H$ is much larger than
the other three ones. As a result, the state $\bar \nu_s$ evolves independently of the others.
Extracting from  $\bar H$ the submatrix with indices  $(1,2,3)$, one obtains the $3\times 3$ 
Hamiltonian
% ------------------------------------------------------------------------------------------------------------------------------------
\begin{equation} \label{eq:Hbar_3nu}
   \bar H_{3\nu} = \bar H_{3\nu}^{\rm {kin}} + \bar H_{3\nu}^{\rm {dyn}}\,
 \end{equation}
% ------------------------------------------------------------------------------------------------------------------------------------
governing the evolution of the $(\bar \nu_{e}, \bar \nu_{\mu}, \bar \nu_{\tau})$ system,
whose dynamical part has the form 
% ------------------------------------------------------------------------------------------------------------------------------------
\begin{eqnarray} \label{eq:Hdyn_1}
\footnotesize
\arraycolsep=3pt
\medmuskip = 1mu
    \bar H^{\rm {dyn}} = 
   V_{CC}   \,  \begin{pmatrix}
	|\bar U_{e1}|^2 + r|\bar{U}_{s1}|^2 & r\bar{U}_{s1}^* \bar{U}_{s2} & r\bar{U}_{s1}^* \bar{U}_{s3}
	\\
	\dagger & r|\tilde{U}_{s2}|^2
	& r\bar{U}_{s2}^* \bar{U}_{s3}
	\\
	\dagger & \dagger
	& r|\bar{U}_{s3}|^2
    \end{pmatrix} \,,
\end{eqnarray}
% ------------------------------------------------------------------------------------------------------------------------------------
where, for brevity, we have indicated with $\dagger$ the complex conjugate of the matrix
element having the same two indices inverted. In deriving Eq.~(\ref{eq:Hdyn_1}) we have made use of the
relations $\bar U_{e2} = \bar U_{e3} = \bar U_{\mu 3} = 0$. Considering the explicit expressions
of the elements of $\bar U$, Eq.~(\ref{eq:Hdyn_1}) takes the form
% ------------------------------------------------------------------------------------------------------------------------------------
\begin{eqnarray} \label{eq:Hdyn_2}
\footnotesize
\arraycolsep=3pt
\medmuskip = 1mu
    \! \bar H^{\rm{dyn}} \!\! = \! 
   V_{\rm{CC}}    \! \begin{pmatrix}
	c_{14}^2 + r c_{34}^2 c_{24}^2 s_{14}^2 \!\!& r c_{34}^2 c_{24} \tilde s_{14} s_{24} \!\! & r c_{34} c_{24} \tilde s_{14} \tilde s_{34}^*
	\\
	\dagger & r c_{34}^2 s_{24}^2 & r c_{34} s_{24} \tilde s_{34}^* 
	\\
	\dagger & \dagger
	& r s_{34}^2
    \end{pmatrix}
\end{eqnarray}
% ------------------------------------------------------------------------------------------------------------------------------------
which,  for vanishing sterile neutrino angles ($\theta_{14} = \theta_{24} = \theta_{34} = 0$) 
returns the standard 3-flavor MSW potential. For small values of such mixing angles, the
dynamical term is approximated by
% ------------------------------------------------------------------------------------------------------------------------------------
\begin{eqnarray} \label{eq:Hdyn_2}
\footnotesize
\arraycolsep=3pt
\medmuskip = 1mu
    \bar H^{\rm{dyn}}  \approx  
   V_{\rm{CC}} \!   \begin{pmatrix}
	1 - (1-r) s^2_{14}  & r  \tilde s_{14} s_{24}  & r  \tilde s_{14} \tilde s_{34}^*
	\\
	\dagger & r  s_{24}^2 & r s_{24} \tilde s_{34}^* 
	\\
	\dagger & \dagger
	& r s_{34}^2
    \end{pmatrix} \,.
\end{eqnarray}
% ------------------------------------------------------------------------------------------------------------------------------------
This shows that the corrections to the standard potential are of second order%
%%%%%%%%%%%%%%%%%%%%%%%%%%%%%%%%%%%%%%%%%%
\footnote{A similar behavior has been observed to occur in solar neutrino
transitions induced by sterile species~\cite{Palazzo:2011rj}.}
%%%%%%%%%%%%%%%%%%%%%%%%%%%%%%%%%%%%%%%%%%
in the new $s_{ij}$'s and further suppressed by a factor $r$ or ($1-r$); in the Earth crust,
$r\sim 0.5$. Therefore, for realistic values of the mixing angles, these
corrections are $O(\epsilon^2)$, and hence at the level of a few per cent. Taking into account 
that in T2K the {\em standard} matter effects are also small, being $v = V_{\rm{CC}}/|k_{13}| \sim 0.05$,
we can deduce that the new nonstandard effects have amplitudes of a few per mill and  have a 
completely negligible impact. We have numerically checked that  allowing for
very large values of $\theta_{34}$ (which is the least known of the three new mixing angles)
even beyond the range currently allowed by the global fits, the transition probability is 
basically indistinguishable from the case of $\theta_{34} = 0$. Therefore, the independence of
the transition probability from $\theta_{34}$, which is exact in vacuum,
remains essentially valid also in the presence of matter effects.

For the calculation of the transition probability, it is useful to define the evolution operator,
which, in the rotated basis, takes the form
% ------------------------------------------------------------------------------------------------------------------------------------
\begin{equation}
    \bar {S} \equiv e^{-i \bar H L} \approx
    \begin{pmatrix}
   	e^{-i \bar H_{3\nu} L} & \boldsymbol{0} \\
	\boldsymbol{0} & e^{-i k_{14} L}
    \end{pmatrix}\,,
\end{equation}
% ------------------------------------------------------------------------------------------------------------------------------------
and is connected to the evolution operator in the original flavor basis through the
unitary transformation
% ------------------------------------------------------------------------------------------------------------------------------------
\begin{equation}
     S = \bar{U} \bar{S} \bar{U}^\dagger \,.
\end{equation}
% ------------------------------------------------------------------------------------------------------------------------------------
Taking into account the block-diagonal form of $\bar{S}$ and the relations $\bar U_{e2} = \bar U_{e3} = \bar U_{\mu 3} = 0$, one has for the relevant transition amplitude
% ------------------------------------------------------------------------------------------------------------------------------------
\begin{eqnarray}
S_{e\mu} = \bar U_{e1} \left[\bar {U}_{\mu 1}^*  \bar {S}_{ee} +  \bar {U}_{\mu 2}^* \bar {S}_{e\mu}\right] 
+ \bar U_{e4} \bar U_{\mu4}^*\bar S_{ss}\,.
\end{eqnarray}
% ------------------------------------------------------------------------------------------------------------------------------------
Since  $\bar S_{ss} =  e^{-i k_{14} L}$ oscillates very fast, 
the associated terms are averaged out by the finite energy resolution of the detector, 
and  for the transition probability we have
% ------------------------------------------------------------------------------------------------------------------------------------
\begin{eqnarray}
\label{eq:pme_gen}
    P_{\mu e}^{4\nu}  \equiv |S_{e\mu}|^2 &=& |\bar U_{e1}|^2 |\bar U_{\mu 1}|^2  |\bar S_{ee}|^2\\
    \nonumber
    &+&   |\bar U_{e1}|^2 |\bar U_{\mu 2}|^2 |\bar S_{e \mu}|^2\\
    \nonumber
      & +&  2 |\bar U_{e1}|^2  {\rm Re} [ \bar U_{\mu 1}^* \bar U_{\mu 2}  \bar S_{ee} \bar S_{e \mu}^*]\\
     \nonumber
      &+&  |\bar U_{e4}|^2 |\bar U_{\mu 4}|^2\,.
\end{eqnarray}
% ------------------------------------------------------------------------------------------------------------------------------------
This expression is completely general, except for the assumption of averaged oscillations,
and connects the 4-flavor transition probability to the amplitudes 
of the effective 3-flavor system governed by the effective Hamiltonian in Eq.~(\ref{eq:Hbar_3nu}). 
Considering the explicit expressions of the elements of the matrix $\bar U$, Eq.~(\ref{eq:pme_gen})
becomes
% ------------------------------------------------------------------------------------------------------------------------------------
\begin{eqnarray}
\label{eq:pme_gen_mix}
    P_{\mu e}^{4\nu} &=&  c_{14}^2 s_{24}^2 s_{14}^2  \bar P_{ee}^{3\nu} \\    
    \nonumber
     &+& c_{14}^2 c_{24}^2  \bar P_{\mu e}^{3\nu} \\
     \nonumber
     &-& 2 c_{14}^2 c_{24} s_{14} s_{24} {\rm Re} (e^{-i \delta_{14}} \bar S_{ee} \bar S_{e \mu}^*)\\
     \nonumber
     &+&  c_{14}^2 s_{14}^2 s_{24}^2 \,,
    \end{eqnarray}
% ------------------------------------------------------------------------------------------------------------------------------------
where $\bar P_{ee}^{3\nu} \equiv  |\bar S_{ee}|^2$ and $\bar P_{\mu e}^{3\nu} \equiv  |\bar S_{e\mu}|^2$. 
Therefore, the  4-flavor problem is reduced to a more familiar 3-flavor one, for which one needs
to calculate the elements $\bar S_{ee}$ and $\bar S_{e\mu}$.

\begin{figure}[t!]
\vspace*{-3.00cm}
\hspace*{-0.50cm}
\includegraphics[width=12.0 cm]{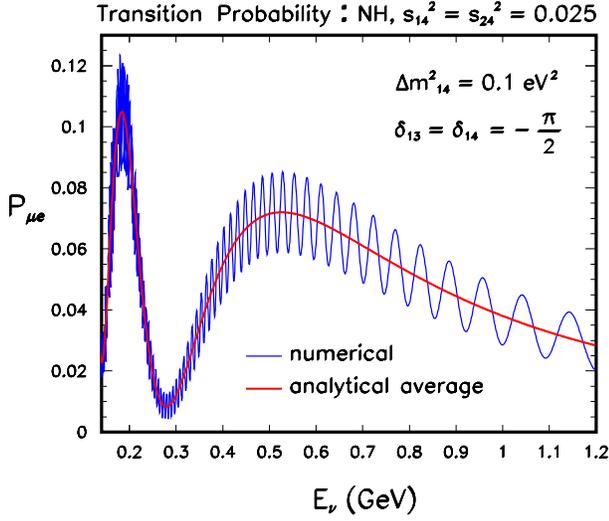}
\vspace*{-2.6cm}
\caption{Probability of $\nu_\mu \to \nu_e$ transition in the 3+1 scheme. The thin blue line 
represents the numerical result, while the red line represents the averaged probability obtained
using Eq.~(\ref{eq:pme_gen_mix}). In both cases the hierarchy is normal and
the  mixing angles are fixed at the values $s_{14}^2 =  s_{24}^2 = 0.025$.
\label{fig:4pan_pme_num}}
\end{figure}

\begin{figure}[b!]
\vspace*{-2.80cm}
\hspace*{-0.50cm}
\includegraphics[width=12.0 cm]{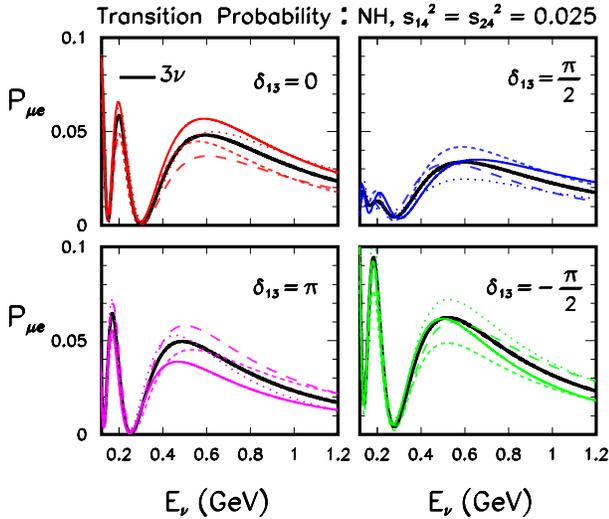}
\vspace*{-2.6cm}
\caption{Probability of $\nu_\mu \to \nu_e$ transition in the 3+1 scheme for normal hierarchy. The four panels
correspond to four different values of the standard {\em CP}-phase $\delta_{13}$. In each panel,
the black thick solid line represents the 3-flavor case ($\theta_{14} = \theta_{24} = 0$),
while the colored lines represent the 4-flavor case (with $s_{14}^2 =  s_{24}^2 = 0.025$) 
for the following four different values of the nonstandard {\em CP}-phase: $\delta_{14} = 0$ (solid), 
$\delta_{14} = \pi$ (long-dashed),  $\delta_{14} = \pi/2$ (short-dashed), and $\delta_{14} = -\pi/2$ (dotted).
\label{fig:4pan_pme}}
\end{figure}

Figure~\ref{fig:4pan_pme_num} shows the fast oscillating transition probability obtained by a
full 4-flavor numerical evolution and the averaged probability calculated using Eq.~(\ref{eq:pme_gen_mix}) 
for the best-fit values $\delta_{13} = \delta_{14} = -\pi/2$. For clarity we have chosen the value $\Delta m^2_{14} = 0.1\,$eV$^2$. 
Figure~\ref{fig:4pan_pme} displays some selected examples of the 4-flavor probability calculated using Eq.~(\ref{eq:pme_gen_mix}).
The four panels correspond to four different values of the standard {\em CP}-phase $\delta_{13}$. In each panel,
the black thick solid line represents the 3-flavor case ($\theta_{14} = \theta_{24} = 0$),
while the colored lines represent the 4-flavor case (with $s_{14}^2 =  s_{24}^2 = 0.025$) 
for four different values of the nonstandard {\em CP}-phase: $\delta_{14} = 0$ (solid), 
$\delta_{14} = \pi$ (long-dashed),  $\delta_{14} = \pi/2$ (short-dashed), and $\delta_{14} = -\pi/2$ (dotted).

%%%%%%%%%%%%%%%%%%%%%%%%%%%%%%%%%%%%%%%%%%%
%%%%%%%%%%%%%%%%%%%%%%%%%%%%%%%%%%%%%%%%%%%%

%%%%%%%%%%%%%%%%%%%%%%%%%%%%%%%%%%%%%%%%%%%
%%%%%%%%%%%%%%%%%%%%%%%%%%%%%%%%%%%%%%%%%%%%

While the 3-flavor elements $\bar S_{ee}$ and $\bar S_{e\mu}$ can be evaluated numerically
(as we have done)  approximate expressions already existing in the literature
in various limits may help to further simplify the expression of the transition probability
in Eq.~(\ref{eq:pme_gen_mix}), which, for small values of the two mixing angles $\theta_{14}$ and  
$\theta_{24}$, takes the form
% ------------------------------------------------------------------------------------------------------------------------------------
\begin{eqnarray}
\label{eq:pme_exp_mix}
    P_{\mu e}^{4\nu} &\simeq& (1 - s_{14}^2 -s_{24}^2) \bar P_{\mu e}^{3\nu} \\
    \nonumber
     &-& 2 s_{14} s_{24} {\rm Re}  (e^{-i \delta_{14}} \bar S_{ee} \bar S_{e \mu}^*)\\
     \nonumber
    &+& s_{14}^2 s_{24}^2  (1+\bar P_{ee}^{3\nu})\,.
  \end{eqnarray}
% ------------------------------------------------------------------------------------------------------------------------------------
First, it can be noted that for small values of $s_{13} \sim \epsilon$ and $\alpha \Delta \sim \epsilon^2$
one has~\cite{Kikuchi:2008vq} 
%------------------------------------------------------------------------------------------------------------------------------------
\begin{eqnarray}
\label{eq:S_exp_ee}
\bar S_{ee} &\simeq& 1 - O(\epsilon ^2)\,.
\end{eqnarray}
% ------------------------------------------------------------------------------------------------------------------------------------
Since we are interested to terms up to  $O(\epsilon^4)$,  we can assume $\bar S_{ee} = 1$.
Moreover, as discussed above, the nonstandard matter effects are completely negligible and 
only the small standard matter effects are relevant. In this approximation, the 3-flavor amplitude $\bar S_{e\mu}$ 
has the well-known (see, for example,~\cite{Kikuchi:2008vq}) form
% ------------------------------------------------------------------------------------------------------------------------------------
\begin{eqnarray}
\label{eq:S_exp_em}
\bar S_{e\mu} &\simeq& A s_{13}^{m} \sin \Delta^m    +  B (\alpha \Delta)\,, 
\end{eqnarray}
% ------------------------------------------------------------------------------------------------------------------------------------
where $A$ and $B$ are two complex coefficients with $O(1)$ modulus, given by
% ------------------------------------------------------------------------------------------------------------------------------------
\begin{eqnarray}
\label{eq:A}
   A &=&  - 2\,  i\, s_{23} e^{-i(\Delta + \delta_{13})} \,, \\   
\label{eq:B}
       B  &=& - 2\, i\, c_{23} s_{12} c_{12} \,,
 \end{eqnarray}
% ------------------------------------------------------------------------------------------------------------------------------------
and ($s_{13}^{m},\Delta^{m}$) are the approximated expressions of ($s_{13}, \Delta$)
in matter
% ------------------------------------------------------------------------------------------------------------------------------------
\begin{eqnarray}
\label{eq:s_d_mat_s13}
   s_{13}^m  &\simeq& (1+v) s_{13}\,, \\   
 \label{eq:s_d_mat_Delta}   
    \Delta^{m}  &\simeq&  (1-v) \Delta  \,,
 \end{eqnarray}
% ------------------------------------------------------------------------------------------------------------------------------------
with $v = V_{\rm {CC}}/|k_{13}| \simeq 0.05$. Making use of Eqs.~(\ref{eq:S_exp_ee})-(\ref{eq:s_d_mat_Delta})
in the expression of the transition probability in Eq.~(\ref{eq:pme_exp_mix}),
in the limit case $v =0$ we recover, in an alternative way, the fourth-order expansion
of the vacuum formula in Eq.~(\ref{eq:Pme_4nu_vac}) presented in Sec.~II. For $v\ne0$, 
one sees that the structure of the transition probability remains the same as in vacuum, 
containing six terms of which three are of the interference type. The only
impact of matter effects (at least for the T2K setup) is to break the degeneracy between 
NH and IH, exactly as it occurs in the 3-flavor case, because of the shifts 
$s_{13} \to s_{13}^m$ and $\Delta \to \Delta^m$ in Eqs.~(\ref{eq:s_d_mat_s13}),(\ref{eq:s_d_mat_Delta}).

\section*{Acknowledgments}
A. P.  acknowledges support from the European Community through a Marie Curie IntraEuropean 
Fellowship, Grant No. PIEF-GA-2011-299582, ``On the Trails of New Neutrino Properties."
We  acknowledge partial support from the European Union FP7 ITN Invisibles (Marie Curie Actions, 
Grant No. PITN-GA-2011-289442).

\bibliographystyle{h-physrev4}

\end{document}